*In-silico* **nucleotide and protein analyses of S-gene region in selected zoonotic coronaviruses reveal conserved domains and evolutionary emergence with trajectory course of viral entry from SARS-CoV2 genomic data**

Adejoke Olukayode Obajuluwa, [Biotechnology Unit, Department of Biological Sciences][1], Pius .Abimbola Okiki[1], Tiwalola Madoc Obajuluwa, [Department of Media and Communication][2]; and Olakunle Bamikole Afolabi [Department of Chemical sciences, Biochemistry Unit][3], Afe Babalola University, Ado Ekiti, Ekiti State, Nigeria.

Corresponding Author: Adejoke. O. Obajuluwa[1] (ibitayoao@abuad.edu.ng), Department of Biological Sciences (Biotechnology Unit), Afe Babalola University.

**ABSTRACT**

The recent zoonotic coronavirus virus outbreak of a novel type [COVID 19] has necessitated the adequate understanding of the evolutionary pathway of zoonotic viruses which adversely affects human populations for therapeutic constructs to combat the pandemic now and in the future. We analyzed conserved domains of the severe acute respiratory coronavirus 2 [SARS-CoV2] for possible targets of viral entry inhibition in host cells, evolutionary relationship of human coronavirus [229E] and zoonotic coronaviruses with SAR-CoV2 as well as evolutionary relationship between selected SARS-CoV 2 genomic data. Conserved domains with antagonistic action on host innate antiviral cellular mechanisms in SARS-CoV 2 include nsp 11, nsp 13 etc. Also, multiple sequence alignments of the spike [S] gene protein of selected candidate zoonotic coronaviruses alongside the S gene protein of the SARs-CoV2 revealed closest evolutionary relationship [95.6%] with pangolin coronaviruses [S] gene. Clades formed between Wuhan SARS-CoV2 phylogeny data and five others suggests viral entry trajectory while revealing genomic and protein SARS CoV 2 data from Philippines as early ancestors. Therefore, phylogeny of SARS-CoV 2 genomic data suggests profiling in diverse populations with and without the outbreak alongside migration history and racial background for mutation tracking and dating of viral subtype divergence which is essential for effective management of present and future zoonotic coronavirus outbreaks.

## INTRODUCTION

Coronaviruses [CoVs] are enveloped viruses with a positive-sense, single-stranded RNA genome belonging to the Coronaviridae family [1]. CoVs are divided into alpha, beta, gamma, and delta groups, and the beta group is further composed of A, B, C, and D subgroups [2]. The virus belongs to the 2B group of the betacoronavirus family, which includes SARS-CoV and Middle East respiratory syndrome coronavirus-MERS-CoV [3].Their entry into respiratory and oesophageal routes accounts for mild to severe acute respiratory syndromes which has led to global epidemics with high morbidity, mortality and immense economic losses in affected human populations [4,5]. Encoded within the 3′ end of the viral genome are the four main structural proteins of coronavirus particles : spike [S], membrane [M], envelope [E], and nucleocapsid [N] [6] as shown in Fig 1 .Phylogenetic analyses of 15 human CoV whole genomes revealed 2019 novel CoV genome shares highest nucleotide sequence identity with SARS-CoV [79.7%] while its two evolutionarily conserved regions [envelope and nucleocapsid proteins] had sequence homology of 96% and 89.6% with same respectively [3]. Hence, the nomenclature for the novel type of the coronavirus outbreak. Surface proteins which sticks out like crown tips [spikes] on coronaviruses binds to host cell receptors- angiotensin converting enzyme 2 [ACE 2] in epithelial cells in hosts. The S1 subunit [N-terminal] of the surface protein facilitates binding to the ACE2 receptor while the S2 subunit [C-terminal] mediates host cell entry through the binding of the viral S protein to human dipeptidylpeptidase 4 [DPP4], marking onset of infection [7, 8]. Interestingly, conserved domains of CoVs have been indicated in literatures as vital entry targets in vaccine and drug development [9, 10].However, growing variability and mutational changes in viruses can cause lack of specificity and reduce efficiency of therapeutic measures. Recombination serves central function in virus replication and evolution in viral infections such as HIV, Ebola, and MERS [11,12] while molecular mechanisms[*RNA fragmentation and trans-esterification reactions*] are possible causes of RNA fragments ligation and subsequent increased novel recombination frequency observed among various RNA viruses [13]. Diverse host factors accounts for a great deal of genome variability in viral recombinants which ranges from multi-resistance to evolutionary novelties [14]. The emergence of novel viral variants trafficked by humans and animals alike through global travel has remained a constant threat in public health and increasing complexity of host-viral interactivity in viral adaptation and evolution [15 ].

**Methods and materials**

***Comparison and analyses of conserved domain of 2019-nCoV/SARS-CoV-2 protein***

Reference number [initial entry with refSeq number NC_045512.1] SARS-CoV2 was retrieved from NBCI database and query for its conserved domains was launched using CDS resources. Proteins with similar conserved domains were included in the subsequent multiple sequence alignment of spike gene of zoonotic coronaviruses investigated in this study.

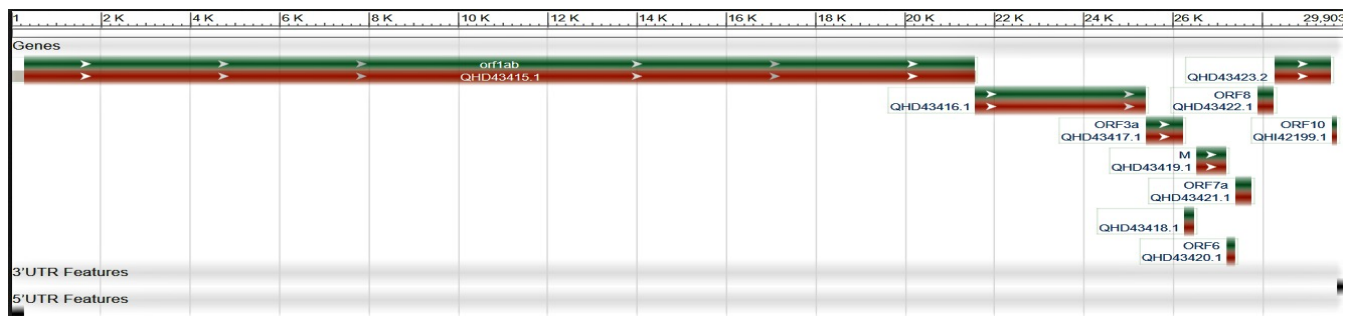

Fig 1: The S, M, N and E regions of the MN908947 genomic data

*Building the spike [S] gene protein candidates from zoonotic coronavirus hosts*

The identification of highly related contigs in a set of viral genomes from Blast search of SARS-CoV like sequences in NCBI database [Supplementary File 1] directed our subsequent search of spike proteins in the selected nine zoonotic coronaviruses. The CoV–host spike proteins that were compared with SARS-CoV 2[QHD43416.1] are : Infectious bronchitis virus-IBV [ADR51590.1], HCoV-229E [AII82124.1], Transmissible gastroenteritis virus-TGEV[AAQ02624.1], Feline infectious peritonitis virus-FIPV[AZH81408.1], Porcine epidemic diarrhea virus-PEDV[AKP16765.2], Equine coronavirus [BAJ52885.1], Murine hepatitis virus-MHV [BAA11889.1], Bovine coronavirus [CCE89341.1], Pangolin coronavirus[QIQ54048.1], Bat coronavirus [AWV67072.1]. Their nucleotide and S gene protein sequences were pooled using NCBI resource tools while analysis was done using EMBOSS Needle, Clustal W2 and Clustal Omega respectively.

**Homology and Phylogeny Analysis of the S-protein genes in candidate zoonotic viruses**

The identified spike gene protein sequences of animal coronaviruses were retrieved from submitted protein entries in NCBI database, homology analysis of the sequences was compared using Clustal Omega, EMBOSS Needle [https://www.ebi.ac.uk/Tools/services/] while phylogenetic trees was constructed using the Neighbor-Joining method by CLUSTAL X software.

*SARS-CoV2 sequence and phylogenetic analyses*

In total, we collected the respective genomic and protein data for eight [8] 2019-nCoV/SARS-CoV-2 isolates which are: [1] MN908947;QHD43415.1 [China-Wuhan, December 2019], [2] LC522350;BBZ90167.1 [Phillipines-January 2020] [3] LC523807;BCA37476.1 [Philipines-January 2020], [4]LC523808;BCA37477.1[Philipines-January 2020]; [5]MT308701;QIV64962.1[Tunisia-April 2020] [6] MT308702;QIV64963.1[USA-February

2020], [7] MT308703;QIV64975.1[USA-April 2020], and [8] MT308704[USA-April 2020] from the betacoronaviruses database in NCBI. Whole-genome alignment and protein sequence identity calculation were performed using Multiple Sequence Alignment in EMBL-EBI database [https:// www.ebi.ac.uk/] with default parameters in Clustal W2 and Clustal Omega respectively.

**RESULTS AND DISCUSSION**

Four out of 29 domain hits generated from 2019nCoV/SARS-CoV 2 CDD query were selected based on the E-value scores [Table 1] . These are: non-structural protein [nsp 11], Coronavirus RPolN terminus, non-structural protein [nsp 13], and Corona S2 super family. The region of 2019 nCoV domain which encodes nsp 11 spans from about 18046-19824bp.It was indicated in countering host innate antiviral response via inhibition of type I interferon [IFN] production using NendoU activity-dependent mechanisms in porcine reproductive syndrome viruses [16]. The nsp 11 is also associated with pathways such as programmed cell death evasion, mitogen-activated protein kinase signaling, histone-related, cell cycle and DNA replication and the ubiquitin-proteasome through RNA microarray analysis [17,18,19,20] and few nsp 11 inhibitors include papain-like proteinase[plPRO] and 3C-like main protease-3CLpro[21]. Coronavirus RNA-directed RNA polymerase [RdRp] terminus covers the N-terminal region of the coronavirus. It spans from about 13480-14538bp in SARS-CoV2 and its interaction with nsp3 has been indicated in viral replication especially during early onset of infection [22]. The inhibitors of coronavirus RdRp include ATP inhibitors with mfScores lower than –110 [21]. The nsp 13 is regarded as a highly conserved and multifunctional helicase unit and its spans from about 20662-21537 in the SARS-CoV2 isolate [23]. They are SARS-CoV helicases that are chiefly concerned with RNA processing, DNA replication, recombination and repair, transcription and translation [24]. A few potential inhibitors of nsp13 have been identified [25,26] and they act by interfering with its unwinding and ATPase activities. The Coronavirus S2 super family spans from 23546-25372 and forms the characteristic 'corona' after which the group is named. CoV diversity is reflected in the variable spike proteins [S proteins] and evolves into forms differing in receptor interactions and response to various environmental triggers of virus-cell membrane fusion [27].The C-terminal [S2] domain directs ectodomain fusion of all CoVs spike proteins following receptor binding [28,29]. The level of interactions between the S protein and the virus receptor controls the host cell range [30]. A study showed a switch of species specificity via a mutant mouse hepatitis virus[MHV] construct which conferred horizontal gene transfer and ability to infect feline cells which were initially absent in wild MHV cells [30]. This was achieved via the substitution of the spike glycoprotein ectodomain. Another research [31] also indicated role of natural mutations in reactivity between the receptor binding domain of spike and cross-neutralization between palm civet coronavirus and SARS-CoVs.

**Table 1: Conserved domains in SARS-CoV2 genomic data**

| Domain hits | Accession no | Interval | e-value |
|---|---|---|---|
| NSP11 | pfam06471 | 18046-19824 | 0e+00 |
| Corona_RPol_N | pfam06478 | 13480-14538 | 0e+00 |
| NSP13 | pfam06460 | 20662-21537 | 0e+00 |
| Corona[S2] super family | Cl20218 | 23546-25372 | 0e+00 |

**Protein phylogeny assembly of SARS-CoV2 isolates**

Identification of the origin, natural host [s] and evolutionary pathway of viruses which causes pandemics is essential to understand molecular mechanism of their cross-species interactivity and implementation of a proper control measure [32]. Protein sequence alignment analyses reveals the closest evolutionarily conservation between 2019-nCoV/SARS-CoV-2 and Pangolin S protein with 95.6% similarity and 92.1% identity while 46.8% similarity and 31.2% identity was observed between SARS-CoV-2 and Bat S protein [Supplementary File 2]. This finding therefore agrees with reports indicating pangolin as a more recent ancestor of SARS-CoV2 than bats [33, 34] which could have arisen as a result of recombination [*chimera*] or interactions between Pangolin-CoV-like virus with a Bat-CoV-RaTG13-like virus going by the homology and subclade of SARS-CoV 2 and pangolin S genes from Bat S-gene seen in this study [Fig 2]. Although, some computational analyses prediction of the improbability of direct binding between receptor binding domains (RBDs) in SARS-CoV2 and ACE2 in humans suggests otherwise [35,36], studies have shown demonstrations of cross-species interactivity through structural [*in-silico*], *in-vitro and in-vivo* mechanisms [37-39,31]. Series of *in-vivo* and *in-vitro* RNA recombination leading to vast genetic variability of positive strand RNA viruses has also been reported [13]. Domestication, consumption and wildlife activities which results in natural selection on a human or human-like ACE2 receptor [33, 36] raises the possibility of SARS-CoV 2 emergence from pangolin. The receptor-binding domain [RBD] in the spike protein and functional polybasic [furin] cleavage site at the S1–S2 boundary [40,33] directs viral-host cell interactivity. Earlier reports of "no interactivity" between the S glycoprotein of bat-CoV with human receptor ACE2 [*which is essential for cross-species transmission*] was founded on significantly low levels of sequence similarity [79 to 80%] between bat-CoV and SARS-CoV in genes encoding the viral spike [S] glycoprotein [41]. However, the higher homology score [95.6%] in the receptor S-gene region obtained in this study suggests pangolin as the most likely intermediate host of SARS-CoV 2 than bats with 46.8% score [Supplementary file 2].

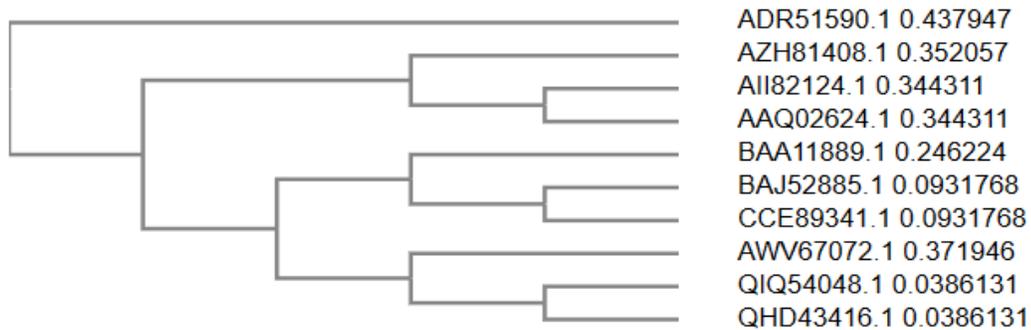

**Fig 2: Spike gene phylogeny of zoonotic CoVs with 2019 nCoV/SARS-CoV 2**

Possible host-viral genetic recombination [42] could account for the increased level of evolutionary divergence observed in submitted entries of the recent SARS-CoV-2 genomic data during time of the study (Supplementary data 2-entries from December 2019 till 4[th] April). Evolutionary patterns observed between Wuhan SARS-CoV 2 data and other five geographical locations reveal trajectory of infection from reported source of outbreak. Mobility patterns of both humans and animals alike are major factors influencing zoonotic disease outbreak amongst populations [43].This amongst others, necessitates the strict travel bans, laws and confinement strategies adopted in different countries to curb its spread. Surprisingly, genomic and protein data from Philipines suggests otherwise (Figs 3 and 4). Despite the limited data used for SARS CoV 2 genomic profiling in this study, we found viral subtype divergence (*considering distance metrics of SARS-CoV2 with entries*) [Figures 3 and 4] suggesting a population-specific post translational modifications which could have been influenced by genetic makeup. This is presumed based on subclades formed between protein sequence data from Philippines [BCA 37476.1 and BCA37477.1] and another between China [QHD 43415.1] and Philippines [BBZ90167.1]-*countries in the same continent* [Figs 4 and 5]. Empirical data points to genetic and epigenetic factors in SARS-CoVs evolution, incidence and infection rates amongst diverse populations and across different racial backgrounds [44].

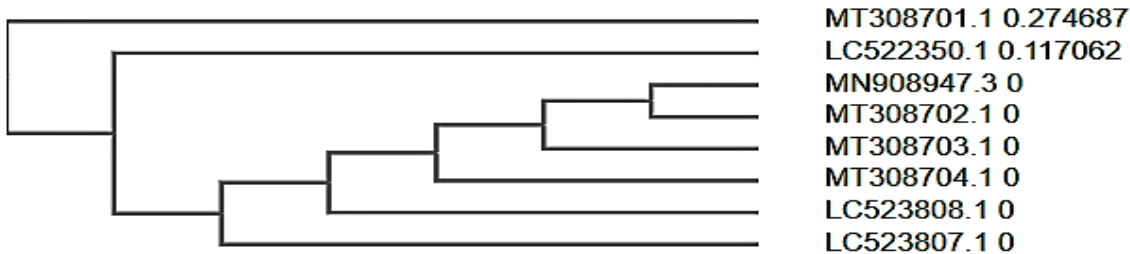

Fig 3: Genomic data phylogeny of selected SARS-CoV 2 entries

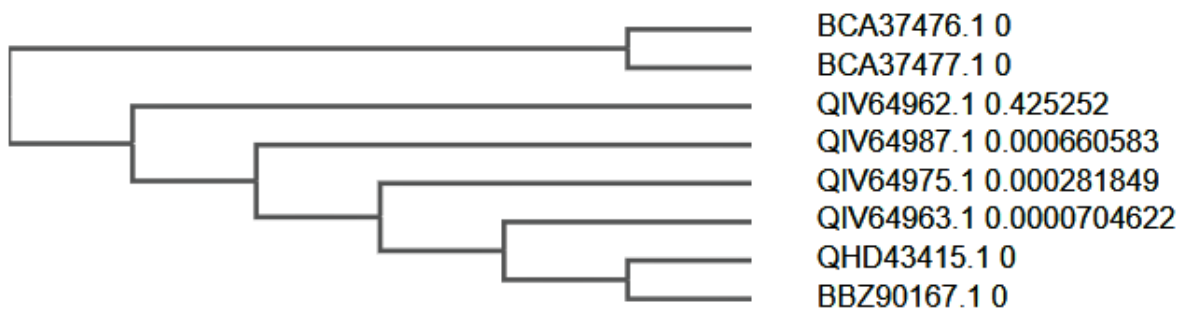

Fig 4: Protein data phylogeny of selected SARS-CoV 2 entries

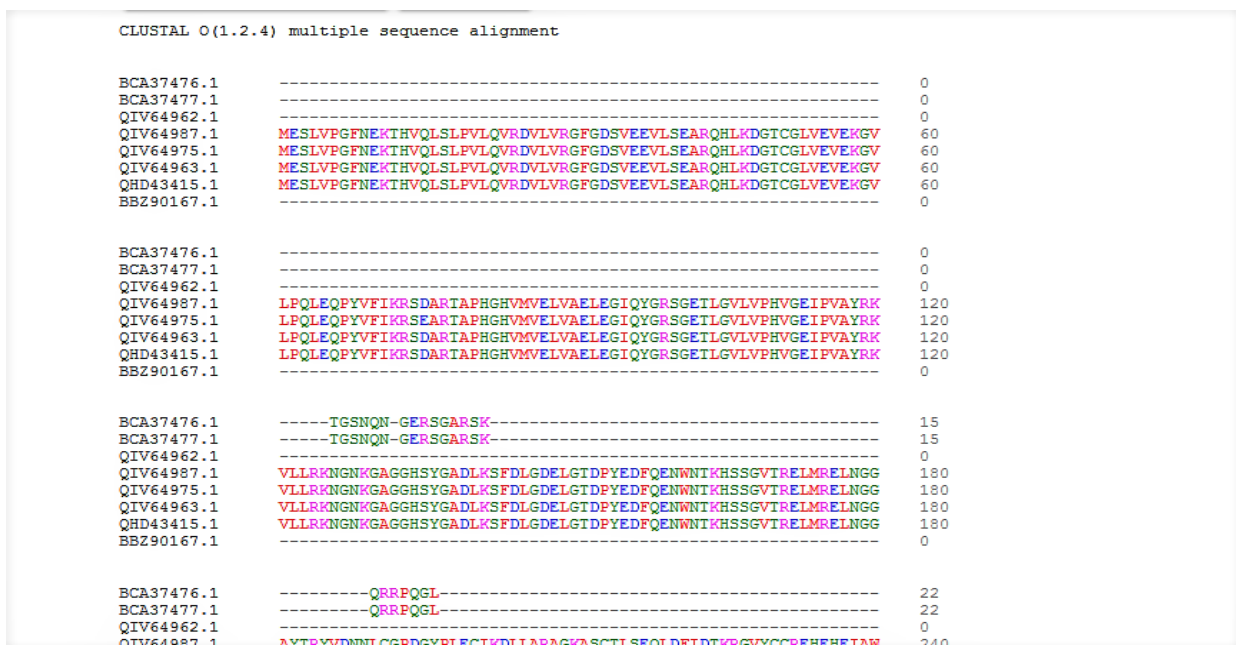

Fig 5: Snapshot of multiple sequence analysis of SARS-CoV 2 protein data

## CONCLUSION

Viral cellular mechanisms are vital factors necessary for replication during infection. Hence, identification of domains of viral entry and evasion of antiviral mechanisms in host cells is essential for development of effective therapeutic measures. Conserved domains that are vital targets sites for inhibition of SARS-CoV2 viral entry and replication in host cells found in this study include nsp11, nsp 13, RdRp and corona super family while compounds such as RNA aptamers, ATP inhibitors, papain-like proteinase [plPRO] and 3C-like main protease-3CLpro etc. are viable indicated inhibitors of these domains. Also, understanding the evolutionary pathway of the novel coronavirus transmission will not only help combat the current pandemic but assist in mutation tracking for identifying future zoonotic coronaviruses threats. The phylogenetic analyses of candidate zoonotic coronavirus [S] gene with SARS-CoV 2 revealed pangolin as the most recent ancestor which formed a subclade with bat S –gene indicating interspecies recombination of CoV in bats and pangolins resulted in SARS-CoV2. Evolutionary pattern observed between SARS-CoV 2 genomic data from source of outbreak with recent entries analyzed in this study showed relative trajectory course of infection from source to other places except protein data from Philippines suggesting earlier existence of SARS-CoV 2 which should be further investigated. Also, genomic and protein data revealed racial viral subtype divergence and rapid rate of mutation despite the novelty of the outbreak. Precise dating of viral subtype divergence will enable researchers correlate divergence with epidemics and pandemics via viral sequence sampling for proper time-scale measurements of zoonotic threats in human populations. Therefore, there is an urgent need for large scale analysis and profiling of genetic data of SARS-CoV2 in affected populations especially in Africa where there is paucity of genomic SARS-CoV data for effective therapeutic measures.

**Conflicts of Interest**

The authors declare no conflict of interest on the work whatsoever.

## Data Availability Statement

The genomic data and multiple sequence analyses results used to support the findings of this study are included within the supplementary information file (s).